\documentclass{aa}
\usepackage{amsmath}
\usepackage{graphicx}
\usepackage[varg]{txfonts}
\usepackage{natbib}
\usepackage{hyperref}
\bibpunct{(}{)}{;}{a}{}{,} 
\begin{document}
\title{Parameters of type IIP SN~2012A and clumpiness effects}

\author{V. P. Utrobin\inst{1,2} \and N. N. Chugai\inst{3}}

\institute{
   Max-Planck-Institut f\"ur Astrophysik,
   Karl-Schwarzschild-Str. 1, 85741 Garching, Germany
\and
   State Scientific Center of the Russian Federation --
   Institute for Theoretical and Experimental Physics of
   National Research Center ``Kurchatov Institute'',
   B.~Cheremushkinskaya St. 25, 117218 Moscow, Russia
\and
   Institute of Astronomy of Russian Academy of Sciences,
   Pyatnitskaya St. 48, 119017 Moscow, Russia}

\date{Received 18 August 2014 / accepted ...}

\abstract{
The explosion energy and the ejecta mass of a type IIP supernova (SN~IIP)
   derived from hydrodynamic simulations are principal parameters of the
   explosion theory. 
However, the number of SNe~IIP studied by hydrodynamic modeling is small.
Moreover, some doubts exist in regard to the reliability of derived SN~IIP
   parameters. 
}{
The well-observed type IIP SN~2012A will be studied via hydrodynamic modeling.
Their early spectra will be checked for a presence of the ejecta clumpiness. 
Other observational effects of clumpiness will be explored.
}{
Supernova parameters are determined by means of the standard hydrodynamic 
   modeling.
The early hydrogen H$\alpha$ and H$\beta$ lines are used for the clumpiness
   diagnostics.
The modified hydrodynamic code is employed to study the clumpiness effect in the
   light curve and expansion kinematics.
}{
We found that SN~20012A is the result of the explosion of a red supergiant 
   with the radius of $715\pm100~R_{\sun}$.
The explosion energy is $(5.25\pm0.6)\times10^{50}$ erg, the ejecta mass is
   $13.1\pm0.7~M_{\sun}$, and the total $^{56}$Ni mass is
   $0.012\pm0.002~M_{\sun}$.
The estimated mass of a progenitor, a main-sequence star, is $15\pm1~M_{\sun}$.
The H$\alpha$ and H$\beta$ lines in early spectra indicate that outer ejecta
   are clumpy.
Hydrodynamic simulations show that the clumpiness modifies the early light curve
   and increases the maximum velocity of the outer layers.
}{
The pre-SN~2012A was a normal red supergiant with the progenitor mass of
   $\approx 15~M_{\sun}$.
The outer layers of ejecta indicate the clumpy structure.
The clumpiness of the external layers can increase the maximum expansion
   velocity.
}
%
\titlerunning{Hydrodynamic model of SN 2012A}
\authorrunning{V. P. Utrobin and N. N. Chugai}
\maketitle

\section{Introduction}
\label{sec:intro}
The general picture of type IIP supernovae (SNe~IIP) phenomenon has been
   understood decades ago \citep{GIN_71, FA_77, EWWP_94}. 
The central to this picture is the explosion of a massive red supergiant (RSG)
   with the energy of the order of $10^{51}$ erg.
Yet the major characteristics -- the explosion mechanism and the progenitor
   mass -- remain the matter of debate.
The hydrodynamic modeling of the well-observed SNe~IIP is the only way to
   determine the explosion energy and the ejecta mass.
The progenitor mass can be obtained via combining the ejecta mass with the
   mass of neutron star and the estimated mass lost by the stellar wind.
In some cases the progenitor mass can be also recovered from the pre-explosion
   images \citep{Sma_09}.
The application of the hydrodynamic approach requires the well-observed SNe~IIP
   with the reliably defined duration of the light curve plateau. 
The number of these well-studied events is rather small: at present only eight
   SNe~IIP are studied hydrodynamically \citep{UC_13}.
Every other well-observed SN~IIP therefore is a bonanza for researchers.

The type IIP SN~2012A in the nearby galaxy NGC 3239 became a subject of the
   detailed observational and theoretical study, including hydrodynamic modeling
   \citep{TCF_13}.
The derived parameters seem reasonable except for the small pre-SN radius.
Moreover, the hydrodynamic model of \citeauthor{TCF_13} produces surprisingly
   low velocity at the photosphere that does not exceed $3000$ km\,s$^{-1}$
   despite the early H$\alpha$ profile shows the expansion velocity up to 
   $\sim10^4$ km\,s$^{-1}$; in addition the Fe\,{\sc ii} 5169 \AA\ absorption
   indicates the photospheric velocity of 5500 km\,s$^{-1}$ on day 15
   \citep{TCF_13}.
We therefore find it appropriate to revisit the hydrodynamic modeling of this
   object.

Another motivation for us to consider SN~2012A originates from the fact that
   the early ($t<20$ d) H$\alpha$ and H$\beta$ lines of type IIP SN~2008in
   reveal a serious problem (we dub it H$\alpha$/H$\beta$ problem): the model
   H$\beta$ absorption turns out too weak compared to the observed absorption,
   the model H$\alpha$ line being consistent with that observed. 
It is remarkable that SN~1987A does not show the H$\alpha$/H$\beta$ problem 
   which prompts a conjecture that this problem is a specific feature of normal
   SNe~IIP originated from the RSG explosion \citep{UC_13}.
The H$\alpha$/H$\beta$ problem can be resolved by invoking the clumpy structure
   of the external ejecta \citep{CU_14}. 
Keeping in mind that SN~2012A in many respects is similar to SN~2008in, it is
   of great interest to explore whether the H$\alpha$/H$\beta$ problem arises
   in the case of SN~2012A as well. 
If confirmed, the next question should be posed: what are the other observational
   effects of the proposed ejecta clumpiness?
This issue could be explored, for example, by means of one-dimensional
   hydrodynamic simulations which somehow take into account the clumpy structure
   of the external ejecta.

Here we perform the hydrodynamic modeling of SN~2012A in order to derive the
   SN parameters using a standard approach applied earlier to other SNe~IIP.
We then analyze the early spectra to check whether the H$\alpha$/H$\beta$
   problem arises for SN~2012A which would indicate the ejecta clumpiness.
As we will see this is indeed the case.
We also explore the issue of observational effects of the clumpiness on
   the basis of a modified one-dimensional hydrodynamic code. 
An additional motivation to study this subject stems from the unsuccessful 
   search for a missing factor responsible for the "mass problem" revealed
   first for SN~2005cs -- the conflict between the high mass obtained from
   the hydrodynamic modeling \citep{UC_08} and the low mass recovered from
   archival images \citep{MSD_05}.
 
\section{Observational data}
\label{sec:obsdat}
The hydrodynamic modeling with the one-group radiation transfer is aimed at
   reproducing a bolometric light curve and photospheric velocities.
For SN~2012A the bolometric light curve is recovered using $UBVRIJHK$ photometry
   measured by \citet{TCF_13} and corrected for the reddening
   $E(B-V)=0.037^{-0.006}_{+0.008}$ mag adopted by them. 
We use a black-body spectral fit to calculate the integrated flux with the
   zero-points reported by \citet{BCP_98}.
Following \citet{TCF_13}, we adopt the distance modulus of $m-M=29.96\pm0.15$
   mag to the nearby galaxy NGC 3239.
We use our hydrodynamic model and the calculated $R$-band light curve together
   with the $R$ magnitude at the SN detection \citep{MNP_12} to fix the
   explosion epoch at MJD=55930.6.
This date is 2.4 days before the explosion moment estimated by \citet{TCF_13}.
Both values however are consistent within the errors.
Below we count time from our explosion date.
The photospheric velocities for several moments are derived by the modeling of
   line profiles.
From the H$\alpha$ and H$\beta$ lines we find the velocity values of 9000, 6100,
   5400, and 1800~km\,s$^{-1}$ on day 5.5, 12.5, 22.4, and 52.4, respectively.
The velocity uncertainty does not exceed $\pm100$ km\,s$^{-1}$. 

\section{Model overview}
\label{sec:mod}

\subsection{Standard hydrodynamic model}
\label{sec:mod-hyd}
The numerical modeling of a SN outburst exploits the implicit, Lagrangian,
   radiation hydrodynamics code {\sc Crab} which integrates the
   spherically-symmetric hydrodynamic equations with a gravity force and
   radiation transfer equation in the one-group (grey) approximation
   \citep{Utr_04, Utr_07}.
The one-group radiation transfer of the {\sc Crab} code is rather accurate
   approximation for the problems we deal with which is supported by the
   comparison of the parameters of type IIP SN~1999em recovered by
   \citet{Utr_07} with those obtained by \citet{BBP_05} in the framework
   of their multi-group radiation hydrodynamics code {\sc Stella}.

SN~2012A resembles photometrically and spectroscopically the type IIP SN~2008in
   \citep{RKB_11} which suggests that parameters of these SNe are close and
   that in the case of SN~2012A pre-SN is also a RSG star.
We will use a non-evolutionary RSG model in the hydrostatic equilibrium for the
   pre-SN which is exploded by a supersonic piston applied to the bottom of
   the stellar envelope at the boundary with the $1.4~M_{\sun}$ central core.
The core presumably collapses into a neutron star and remains outside
   the computational domain.

\subsection{Modification for clumpy ejecta}
\label{sec:mod-clump}
In order to explore the effects of the clumpy structure of the outer layers, we
   modify {\sc Crab} code by means of introducing the clumpiness only in the
   radiation transfer equation leaving the hydrodynamics intact.
However, the clumpiness affects the hydrodynamics implicitly, since the radiative
   force is modified by the clumpiness via the radiation transfer effects of
   the clumpy medium.

\subsubsection{Radiative transfer in clumpy medium}
\label{sec:mod-clump-radtr}
The radiation transfer in the clumpy medium is treated by using the standard 
   equations in which the absorption (or scattering) coefficient and the
   emissivity are modified by the inclusion of the clumpiness.
We consider a clumpy structure of the outer layers as a medium composed by an 
   ensemble of the dense clumps of a density $\rho_c$ embedded in a more tenuous
   interclump medium of a density $\rho_i$.
With the clump-to-average density contrast $\chi=\rho_c/\rho$ and the mass
   fraction of clumps $\mu$, the volume filling factor of clumpy component is
   $f=\mu \chi^{-1}$, while the clump and interclump densities are 
\begin{equation}
 \rho_c = \chi \rho \quad\mbox{and}\quad \rho_i = \frac{1-\mu}{1-f} \rho\; .
 \label{eq:dens}
\end{equation}
We assume that clumps are uniform spheres of a radius $a$ which are randomly
   distributed but do not overlap.
The number density of clumps is then
\begin{equation}
   n_c = \frac{3 f}{4 \pi a^3} \; .
   \label{eq:nmbden}
\end{equation}
A random photon traveling a length $s$ shares its path between the clumps,
   $fs$, and the interclump medium, $(1-f)s$ \citep{KM_63}. 
This suggests that the absorption coefficient in a clumpy medium can be written
   as a sum
\begin{equation}
   k_{tot} = f k_c^{eff} + (1 - f) k_i \; ,
   \label{eq:ktot}
\end{equation}
where $k_c^{eff}$ is the effective absorption coefficient for the clumps and
   $k_i$ is the absorption coefficient for the interclump medium. 
The absorption coefficient $f k_c^{eff}$ of the clumpy component enters the
   element of the optical depth $d\tau_{cl}$ along the linear displacement $ds$  
\begin{equation}
  d\tau_{cl} = f k_c^{eff} ds = \pi a^2 n_c p ds \; ,
   \label{eq:dtau}
\end{equation}
   where $p$ is the average absorption probability for the photon randomly
   striking the cloud.
After the elementary integration \citep[e.g.,][]{HP_93}, the absorption 
   probability reads 
\begin{equation}
   p = 
   1 - \frac{1}{2 \tau_c^2} + \left(\frac{1}{\tau_c} + 
   \frac{1}{2 \tau_c^2}\right) \rm{e}^{-2\tau_c} \; ,
   \label{eq:avrp}
\end{equation}
where $\tau_c = k_c a$ is the clump optical thickness, $k_c$ is the
   microscopic absorption coefficient of the clump matter.
As expected, the absorption probability $p=1$ for the $\tau_c \gg 1$, 
   and $p=4\tau_c/3$ for $\tau_c \ll 1$.
Introducing the value $q(\tau_c)=(3/4\tau_c)p(\tau_c)$ reduces the effective
   absorption coefficient $k_c^{eff}$ to
\begin{equation}
   k_c^{eff} = k_c q(\tau_c) \; .
   \label{eq:kceff}
\end{equation}

The emissivity of the clumpy medium is treated in the same way as the absorption
   coefficient, {\it viz.},
\begin{equation}
   \eta_{tot} = f \eta_c^{eff} + (1 - f) \eta_i \; .
   \label{eq:etatot}
\end{equation}
The emissivity of the clumpy component is
\begin{equation}
   f \eta_c^{eff} = \frac{1}{4 \pi} n_c L_c\; ,
   \label{eq:emsfrc}
\end{equation}
   where $L_c = 4 \pi a^2 F$ is the luminosity of a clump and $F$ is the
   radiation flux escaping the clump surface.
Assuming a homogeneous emissivity $\eta_c$ across the clump, the emergent
  intensity in the direction at the angle $\theta$ to the outward normal is
\begin{equation}
   I(\theta) = \int_{0}^{2 \tau_c \cos \theta} \!\!\! \eta_c 
   \exp(-k_c s) \; ds
   = \frac{\eta_c}{k_c} [1-\exp(-2 \tau_c \cos \theta)] \; .
   \label{eq:intemg}
\end{equation}
Integrating the projection $I\cos{\theta}$ over angles gives the flux escaping
   the spherical clump 
\begin{equation}
   F = \frac{4}{3} a \eta_c q(\tau_c) \; .
   \label{eq:flxemg}
\end{equation}
The effective emissivity is thus reduced to 
\begin{equation}
   \eta_c^{eff} = \eta_c q(\tau_c) \; .
   \label{eq:etaeff}
\end{equation}
Remarkably, the expressions for the effective absorption coefficient
   (\ref{eq:kceff}) and the effective emissivity (\ref{eq:etaeff}) look
   similar; this is an outcome of the optical reversibility. 
It is worth to note that the function $q(\tau_c)$ can be interpreted as 
   the escape probability for a photon emitted in a spherical homogeneous clump
   \citep{Ost_89}.

\subsubsection{Setting out clumpiness}
\label{sec:mod-clump-hydclmp}
The ejecta clumpiness suggested earlier for SN~2008in \citep{UC_13, CU_14} is
   presumably generated during the shock wave propagation in the outermost layers
   of a RSG star which are characterized by the presence of a density inversion
   and convection \citep[e.g.,][]{Pac_69, Mae_81, CFMP_11}. 
The mass of these layers depends on the RSG mass and mounts to
   $0.01-0.2 M_{\sun}$ for the stellar mass in the range of $10-20~M_{\sun}$
   \citep{Fad_12}.
Two mechanisms could be involved in the clumpiness production.
The first is related to the shock wave propagation through the density inversion
   layer which should result in the Rayleigh-Taylor and Richtmyer-Meshkov
   instabilities.
The second mechanism is related to the shock wave propagation through the
   outer convective zone.
The convection velocity in a RSG probed by the macroturbulent velocity attains
   $6-10$ km\,s$^{-1}$ \citep{CFMP_11}, i.e., comparable to the sound speed. 
The colliding tangential flows of the neighboring convective cells will produce
   supersonic collision accompanied by a significant compression.
This suggests that the convective zone contains the density perturbations of
   large amplitude, $\delta \rho/\rho \sim 1$.
The SN shock wave running through the inhomogeneous convection layer can
   produce the clumpy post-shock flow with the large density contrast.

We set the inhomogeneous structure of the ejecta by turning on the clumpiness
   generation when the shock wave reaches the level corresponding to a certain
   overlying mass, e.g., $\approx 0.07~M_{\sun}$.
The downstream clumpiness parameters $\mu$ and $\chi$ are set to grow with the
   local hydrodynamic time scale from zero to their final values that are
   limited by the pre-set model values $\mu_0$ and $\chi_0$, respectively.
The third parameter, the clump radius $a$, is assumed to be the constant
   fraction of the shell radius $a/r = 0.016$; the value is adopted following
   the estimate on the basis of the amplitude of flux fluctuations in the
   H$\alpha$ line profile of SN~2008in \citep{CU_14}.
 
To facilitate the calculation of temperatures of the clumps and the interclump 
   medium, we assume that they are in a pressure equilibrium.
In the optically thick medium the total pressure of gas and radiation is
   determined by the thermodynamic equilibrium.
In contrast, in the optically thin case the pressure equilibrium is
   predominantly controlled by the gas pressure because of a negligible
   interaction between gas and radiation field and of the same radiation field
   in both the clumps and the interclump medium.
To describe these extreme regimes, we introduce the effective pressure
\begin{equation}
   P_{eff}(\,\rho, T) =
       \begin{cases}
       P_g(\,\rho, T, T_r)& \text{for $\tau \ll 1$\; ,}\\
       P_g(\,\rho, T) + {1\over3} a T^4& \text{for $\tau \gg 1$\; } \; .
       \end{cases}
   \label{eq:peff}
\end{equation}
The intermediate regimes between the optically thick and thin cases are
   described by the factor $\exp(-\tau)$ in the radiation pressure where 
   $\tau$ is the total optical depth at the certain layer from the outer
   boundary of the envelope.

The hydrodynamic code works with a smooth medium described by the density $\rho$
   and the gas temperature $T$ that specify the effective pressure
   $P_{eff}(\,\rho, T)$.
The pressure equilibrium between the clumps and the interclump matter along with
   the effective pressure (\ref{eq:peff}) permits us to calculate the gas
   temperature of the clumps ($T_c$) and interclump matter ($T_i$) from
   the equalities:
\begin{equation}
    P_{eff}(\,\rho, T) = P_{eff}(\,\rho_c, T_c) = P_{eff}(\,\rho_i, T_i) \; .
   \label{eq:tcti}
\end{equation}
The temperatures $T_c$ and $T_i$ combined with the corresponding densities
   $\rho_c$ and $\rho_i$ (\ref{eq:dens}) and the radiation temperature $T_r$
   are used to calculate the absorption coefficients according to relation
   (\ref{eq:ktot}) and the total emissivity (\ref{eq:etatot}) for the clumpy
   medium which enter the radiation hydrodynamics equations.

\begin{figure}[t]
   \includegraphics[width=\hsize, clip, trim=12 375 67 206]{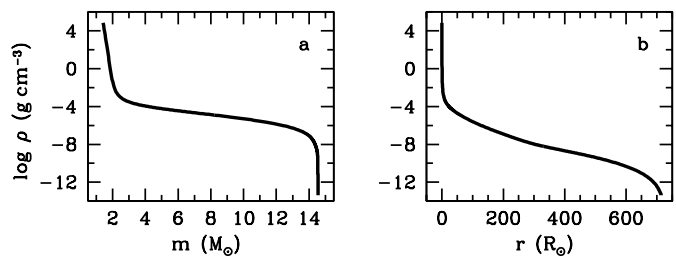}
   \caption{%
   Density distribution as a function of interior mass (Panel \textbf{a}) and
   radius (Panel \textbf{b}) for the optimal pre-SN model of SN~2012A.
   The central core of 1.4 $M_{\sun}$ is omitted.
   }
   \label{fig:denmr}
\end{figure}
\begin{figure}[t]
   \includegraphics[width=\hsize, clip, trim=18 152 67 316]{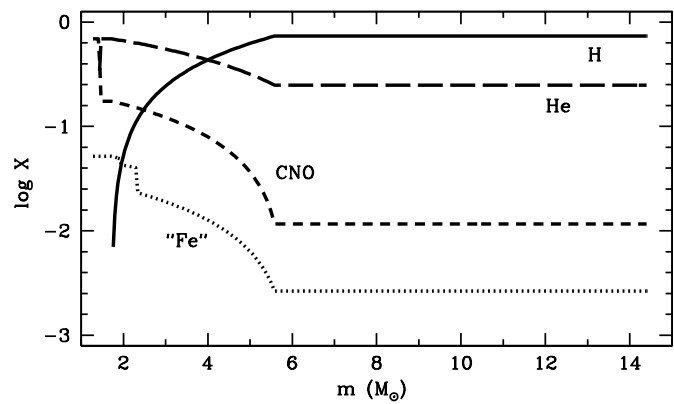}
   \caption{%
   The mass fraction of hydrogen (\emph{solid line\/}), helium
      (\emph{long dashed line\/}), CNO elements (\emph{short dashed line\/}),
      and Fe-peak elements including radioactive $^{56}$Ni
      (\emph{dotted line\/}) in the ejecta of the optimal model.
   }
   \label{fig:chcom}
\end{figure}
%
\section{Supernova parameters}
\label{sec:snpar}
The parameters of SN~2012A are determined in a standard way by means of
   the hydrodynamic simulations of the bolometric light curve and the evolution
   of the photospheric velocity and their fitting to the observations. 
The major model parameters are the ejecta mass, the explosion energy, the pre-SN
   radius, and the total $^{56}$Ni mass.
The latter value is fixed by the bolometric luminosity at the radioactive tail
   and in the case of SN~2012A is equal to $0.012~M_{\sun}$.
Additional tuning parameters are the density distribution in the RSG envelope
   (Fig.~\ref{fig:denmr}), the mixing between the helium core and the
   hydrogen envelope, and the mixing of CNO elements and heavier metals
   dubbed "Fe" elements which also include the radioactive $^{56}$Ni
   (Fig.~\ref{fig:chcom}).
Dependence of the observational properties of a SN~IIP outburst on the
   parameter variations was studied in detail elsewhere \citep{Utr_07}. 
Results of the hydrodynamic modeling are almost independent of the helium-core
   mass, which is taken in accord with the standard evolutionary models for
   a single star of a given initial mass.
In the case of SN~2012A the helium-core mass is adopted to be $4~M_{\sun}$
   which corresponds to the non-rotating star with a ZAMS mass of about
   $15~M_{\sun}$ \citep{HMM_04}.
\begin{figure}[t]
   \includegraphics[width=\hsize, clip, trim=10 155 67 200]{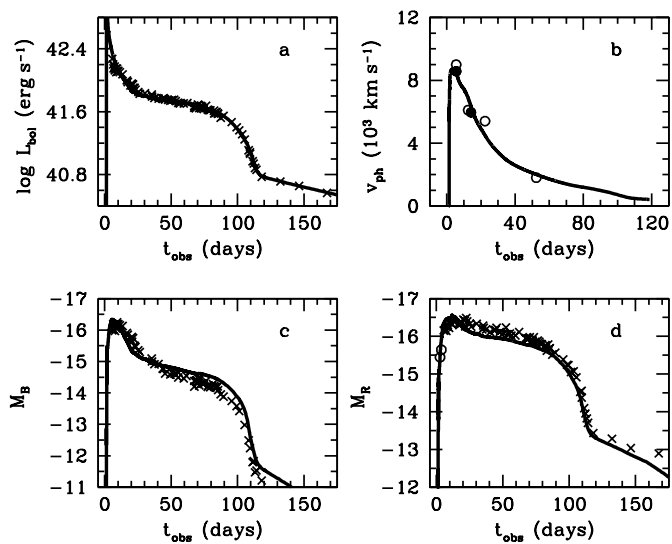}
   \caption{%
   Optimal hydrodynamic model.
   Panel \textbf{a}: the bolometric light curve of the optimal model
      (\emph{solid line\/}) overplotted on the bolometric data of SN 2012A
      (\emph{crosses\/}) evaluated from the $UBVRIJHK$ magnitudes
      reported by \citet{TCF_13}.
   Panel \textbf{b}: the calculated photospheric velocity (\emph{solid
      line\/}) is compared to the photospheric velocity derived
      from the H$\alpha$ and H$\beta$ lines (\emph{open circles\/})
      (Sect.~\ref{sec:obsdat}) and from the He\,{\sc i}~5876~\AA\ line
      (\emph{filled circles\/}) \citep{TCF_13}.
   Panels \textbf{c} and \textbf{d}: the calculated B and R light curves
      (\emph{solid line\/}) compared to the observations of SN 2012A
      (\emph{crosses\/}) obtained by \citeauthor{TCF_13}
   \emph{Two open circles} are the SN detection in $R$-band \citep{MNP_12}.
   }
   \label{fig:blcvph}
\end{figure}
\begin{figure}[b]
   \includegraphics[width=\hsize, clip, trim=15 154 17 272]{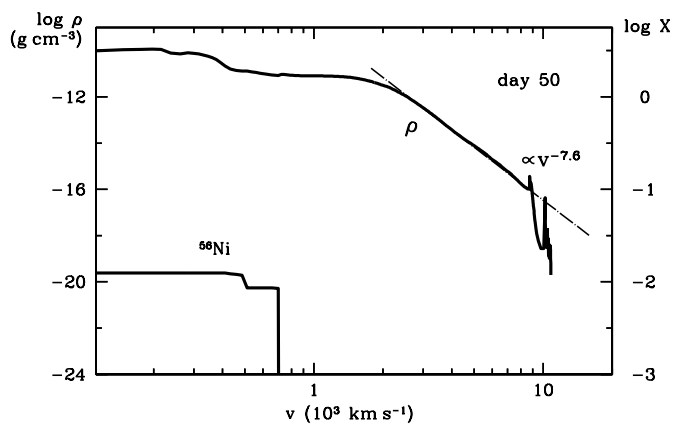}
   \caption{%
   The density and the $^{56}$Ni mass fraction as a function of velocity
      for the optimal model at $t=50$ days (\emph{solid lines\/}).
   \emph{Dash-dotted line\/} is the density distribution fit
      $\rho\propto v^{-7.6}$.
   }
   \label{fig:denni}
\end{figure}
\begin{figure}[t]
   \includegraphics[width=\hsize, clip, trim=37 161 41 252]{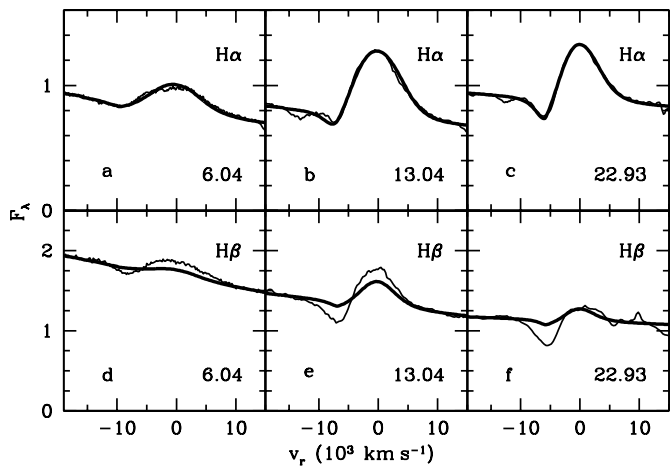}
   \caption{%
   The observed H$\alpha$ (upper panels) and H$\beta$ (lower panels) 
   \citet{TCF_13} in SN~2012A on day 7, 14, and 23 (\emph{thin line\/}) 
    compared to the model profiles (\emph {thick line\/}). The model 
   parameters are adjusted to fit the observed H$\alpha$. For these models, 
   however, the calculated H$\beta$ is unable to fit the observed profile.
   }
   \label{fig:habfit}
\end{figure}

Exploring the parameter space results in the optimal model
   (Fig.~\ref{fig:blcvph}) with the ejecta mass $M_{env}=13.1~M_{\sun}$,
   the explosion energy $E=5.25\times10^{50}$ erg, and the pre-SN radius
   $R_0=715~R_{\sun}$.
The model reproduces not only the bolometric light curve, but also the $B$ and
   $R$-band light curves.
The latter plot is especially valuable, because it demonstrates that the
   earliest photometric points are fitted well and reliably indicate
   the explosion moment. 
The model density distribution in the freely expanding envelope on day 50
   (Fig.~\ref{fig:denni}) is similar to that of SN~2008in \citep{UC_13}
   with the outer density power law $\rho\propto v^{-7.6}$.
The power-law index $k=-\partial \ln \rho / \partial \ln v$ depends on
   the density distribution of pre-SN outer layers, which in turn is
   constrained by the initial luminosity peak. 
The rule of thumb states that a more luminous and longer initial luminosity
   peak requires a shallower density distribution in the outer layers,
   i.e., a lower $k$ value. 
It is worth noting that the modeling of four SNe~IIP, namely, SN~2004et
   \citep{UC_09}, SN~2005cs \citep{UC_08}, SN~2008in \citep{UC_13}, and
   SN~2012A, results in a similar density gradient with $k\approx 7.6$
   in the outer layers.

Combining the ejecta mass with the mass of the neutron star gives the pre-SN
   mass of $14.5~M_{\sun}$.
The progenitor ZAMS mass should be larger by the amount lost by the stellar
   wind.
Following the estimate for SN~2003Z with a comparable progenitor mass
   \citep{UCP_07}, we adopt for SN~2012A the lost mass in the range of
   $0.2-0.8~M_{\sun}$ in which case the progenitor mass turns out to be
   $M=15.0\pm0.3~M_{\sun}$.

The parameter errors can be estimated by varying the model parameters around
   the optimal model. 
Adopting the uncertainty of 17\% in the bolometric luminosity, 4\% in the
   photospheric velocity, and 3\% in the plateau duration, we find the errors
   $\pm100~R_{\sun}$ for the initial radius, $\pm0.7~M_{\sun}$ for the ejecta
   mass, $\pm0.6\times10^{50}$ erg for the explosion energy, and
   $\pm0.002~M_{\sun}$ for the $^{56}$Ni mass.
The error of the ejecta mass combined with the uncertainty in the mass loss
   suggests the progenitor mass error of $\pm1~M_{\sun}$.

\section{Clumpiness effects}
\label{sec:clump}

\subsection{Evidence from hydrogen lines}
\label{sec:clump-hahb}
%
\begin{figure}[t]
   \includegraphics[width=\hsize, clip, trim=53 386 56 190]{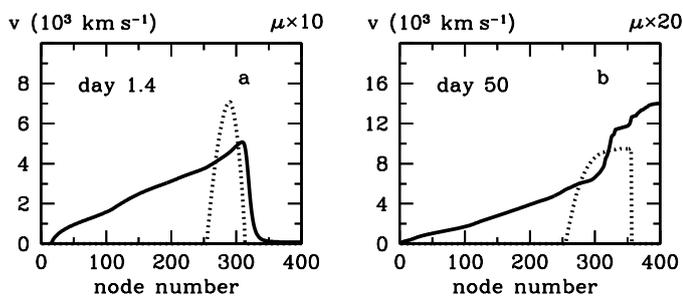}
   \caption{%
   Evolution of velocity and clumping in hydrodynamic model ($\mu_0=0.95$,
      $\chi_0=7$) from the moment just before the shock breakout
      (Panel \textbf{a}) till day 50 (Panel \textbf{b}).
   \emph{Thick line} is the velocity profile and \emph {dotted line} shows
      the mass fraction of clumps.
   Note that the clumpiness is turned off in the outermost layers due to the
      shock breakout.
   }
   \label{fig:devclp}
\end{figure}
\begin{figure}[t]
   \includegraphics[width=\hsize, clip, trim=31 175 40 200]{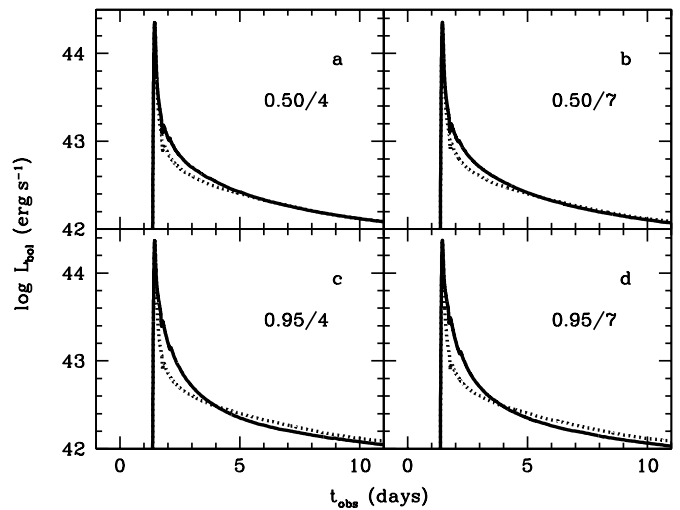}
   \caption{%
   Dependence of the bolometric luminosity (\emph {thick line\/}) at the initial
      peak on the mass fraction of clumps $\mu_0$ and the density contrast
      $\chi_0$ indicated on each panel.
   The bolometric luminosity peak of the optimal model for the smooth medium
      is shown by \emph{dotted line}.
   }
   \label{fig:deplum}
\end{figure}
\begin{figure}[t]
   \includegraphics[width=\hsize, clip, trim=31 175 40 194]{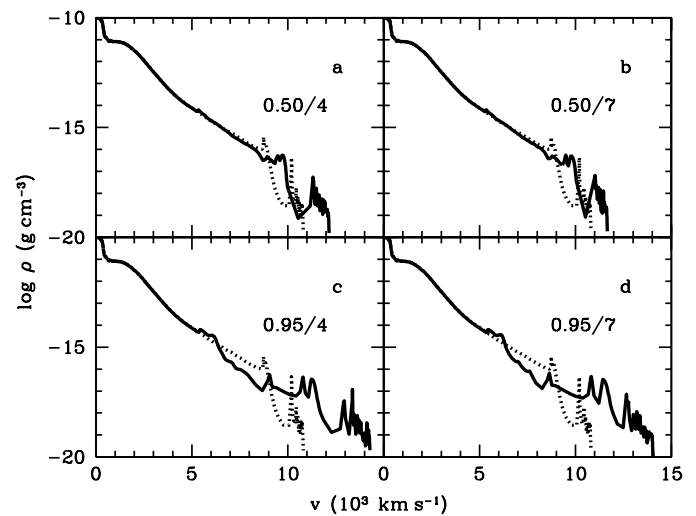}
   \caption{%
   Density as a function of velocity (\emph {thick line\/}) for the
      different mass fraction of clumps $\mu_0$ and density contrast
      $\chi_0$ indicated on each panel.
   Density profile of the optimal model for the smooth medium is shown by
      \emph{dotted line}.
   }
   \label{fig:denv}
\end{figure}
\begin{figure}[t]
   \includegraphics[width=\hsize, clip, trim=31 175 40 194]{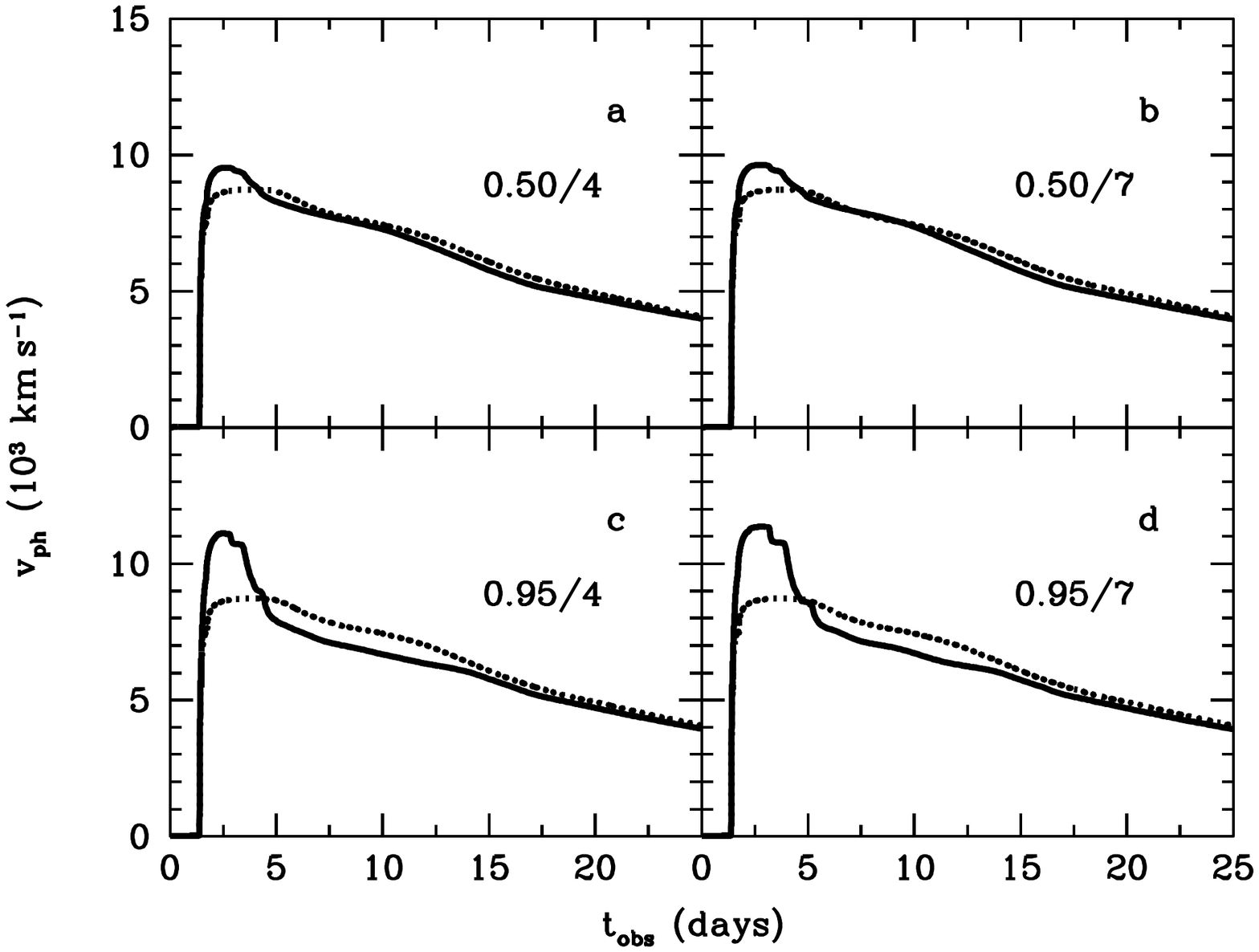}
   \caption{%
   Photospheric velocity evolution (\emph {thick line\/}) for the
      different mass fraction of clumps $\mu_0$ and density contrast
      $\chi_0$ indicated on each panel.
   Photospheric velocity of the optimal model for the smooth medium
      is shown by \emph{dotted line}.
   }
   \label{fig:depvph}
\end{figure}

Recently it was found that the H$\alpha$ and H$\beta$ lines in the early
   ($t \lesssim 20$ d) spectra of SN~2008in cannot be reproduced for the 
   standard spherically-symmetric model \citep{UC_13}; the controversy 
   was resolved by assuming the clumpy structure of the outer layers of the
   ejecta.
To check whether the early SN~2012A spectra reveal the similar
   H$\alpha$/H$\beta$ problem, we apply the standard approach to the description
   of line profiles in an expanding envelope.
We consider a freely expanding atmosphere on the top of the photosphere.
The H$\alpha$ and H$\beta$ line profiles are mainly determined then by the
   radial distribution of the population of the second hydrogen level, $n_2(v)$,
   and the line source function $S(v)$.
The latter consists of the scattering and net emission, $S=W+S_e$, where $W$
   is the dilution factor and $S_e$ is the term responsible for the net emission
   normalized on the photospheric brightness $I$.
In the framework of this model it is easy to fit the H$\alpha$ line on day 7,
   14, and 23 (Fig.~\ref{fig:habfit}).
However, the H$\beta$ absorption component for the best fit function $n_2(v)$
   turns out too weak, i.e., the H$\beta$ line suggests the larger population
   of the second level.
If we proceed the opposite way, i.e., first fit the H$\beta$ line and then use
   the found function $n_2(v)$ to calculate the H$\alpha$ line, we would obtain
   the unacceptably strong H$\alpha$ absorption.
This is exactly the problem we met in the case of SN~2008in.

Following the earlier suggestion for SN~2008in, we assume that the clumpiness
   of the external layers of the SN~2012A ejecta is responsible for the peculiar
   H$\beta$ intensity. 
In this picture the SN atmosphere consists of an ensemble of the dense clouds
   with the filling factor $f$ embedded in a rarefied intercloud medium. 
We managed to fit both the H$\alpha$ and H$\beta$ lines with the filling factor
   $f=(v/v_f)^{-2.8}$ for $v>v_f=5500$ km\,s$^{-1}$ and the velocity at the
   photosphere of 8900, 6200, and 5550 km\,s$^{-1}$ on day 7, 14, and 23,
   respectively.
This requires a clumpy structure of the outer $\approx 0.07~M_{\sun}$ of the
   ejecta in the optimal hydrodynamic model. 

The clumpy model invoked in this case suggests the large clump/interclump ratio
   of the H$\alpha$ optical depth.
On day 14 the H$\alpha$ optical depth in clouds at the fiducial velocity of
   8000 km\,s$^{-1}$ is $\tau({\rm H}\alpha) = 80$ while in the interclump gas
   $\tau({\rm H}\alpha) = 0.5$ at the same velocity, i.e., $160$ times lower.
Since the density contrast of the clouds unlikely significantly exceeds
   $\approx 7$, the compression ratio in the adiabatic radiation dominated
   shock, the factor of $\sim 160$ requires explanation.
In this respect we note that in the case when the population rate of the second
   hydrogen level is dominated by the recombination and the depopulation is
   controlled by the Ly$\alpha$ escape one expects that the hydrogen
   concentration $n_2$ is proportional to $n^3$. 
This means that with the clump/interclump density ratio of $5-6$ the required
   ratio of the H$\alpha$ optical depth in these components can be attained.

\subsection{Hydrodynamic model with clumpiness}
\label{sec:clump-hydmod}
The signatures of the clumpy structure of the ejecta indicated by hydrogen lines
   pose questions concerning other observational effects of the clumpiness.
Following the proposed prescription for the inclusion of the clumpiness into
   the hydrodynamic simulations (Sect.~\ref{sec:mod-clump}), we computed
   a number of models based on the optimal homogeneous model.
The clumpy structure is determined by the adopted mass fraction of clumps
   $\mu$ and the density contrast $\chi$ with the fixed ratio $a/r = 0.016$. 
The clumpiness generation is turned on when the shock wave reaches the external
   mass coordinate of $\approx 0.07~M_{\sun}$.
The evolution of the distributions of $\mu$ and $v$ between the shock breakout
   stage (day 1.4) and the free-expansion regime (day 50) is shown in
   Fig.~\ref{fig:devclp} for the hydrodynamic model with $\mu_0=0.95$ and
   $\chi_0=7$. 
It is noteworthy that in the outermost layers the clumpiness generation is
   turned off at the shock breakout because of the shock radiative damping. 
This explains the sharp drop of the $\mu$ value in the outermost layers clearly
   seen on day 50.

We show results for the combinations of the mass fraction of the clumps,
   $\mu_0 = 0.5$ and 0.95, and the density contrast, $\chi_0 =  4$ and 7. 
These $\chi_0$ values correspond to the adiabatic compression factor for the
   matter and radiation dominated regimes, respectively.
The major effect of the clumpiness is a decrease of the optical depth compared
   to the homogeneous case.
This results in the luminosity enhancement during the first several days
   (Fig.~\ref{fig:deplum}). 
The effect is larger for the larger $\mu_0$ and insensitive to the density 
   contrast $\chi_0$ in agreement with the expression (\ref{eq:dens}) for 
   the density of the interclump medium.
As a result of the flux increase the external layers experience a stronger 
   radiative acceleration thus resulting in the larger maximum velocity of
   the ejecta (Fig.~\ref{fig:denv}).
For the mass fraction of the clumps $\mu_0 = 0.95$ the maximum velocity is about
   30\% larger compared to the homogeneous model. 
The density minimum in the range of $10\,000-12\,000$ km\,s$^{-1}$
   separates the main ejecta and the outer $\sim 10^{-4}~M_{\sun}$ shell
   formed due to the shock breakout.

It is noteworthy that in addition to the enhanced flux the homogeneous structure
   of the outermost layers (Fig.~\ref{fig:devclp}) is another crucial factor
   favoring the larger radiative acceleration.
Indeed, if the outermost layers were also clumpy, the radiation-matter
   interaction would not be strong enough to produce the efficient acceleration.
To check this argument we computed the hydrodynamic model in which the clumpy
   structure was set artificially throughout the external layers. 
This model did not show extra acceleration.

Observationally, velocities of the external ejecta could be probed by the blue
   wings of the absorption components of the H$\alpha$, H$\beta$, and
   He\,{\sc i} 5876~\AA\ lines. 
Unfortunately, the H$\alpha$ and He\,{\sc i} 5876~\AA\ absorptions in the 
   first spectrum of SN~2012A on day 7 \citep{TCF_13} are too shallow for
   confident conclusion.
The H$\beta$ absorption indicates the ejecta velocity up to $11\,000-12\,000$
   km\,s$^{-1}$ which is consistent with the velocity of the density minimum
   (Fig.~\ref{fig:denv}) where the absorption intensity significantly drops.  
We estimate that the H$\beta$ optical depth in the density minimum is
   $\sim 10^{-2}$ on day 7, being beyond the detection limit.   
Another manifestation of the larger expansion velocity in the clumpy model 
   could be the larger photospheric velocity at the early ($t<5$ d) stage 
   compared to the homogeneous model (Fig.~\ref{fig:depvph}). 
Unfortunately, this cannot be confirmed, because the spectra of SN~2012A are
   not available at that early stage.
As to the later ($t>5$ d) stage, the photospheric velocities of smooth and
   clumpy models are close each other and consistent with the observational
   data.

\section{Discussion and Conclusions}
\label{sec:disc}
%
\begin{table}[b]
\caption[]{Hydrodynamic models of type IIP supernovae.}
\label{tab:sumtab}
\centering
\begin{tabular}{@{ } l  c  c @{ } c @{ } c @{ } c  c @{ }}
\hline\hline
\noalign{\smallskip}
 SN & $R_0$ & $M_{env}$ & $E$ & $M_{\mathrm{Ni}}$ 
       & $v_{\mathrm{Ni}}^{max}$ & $v_{\mathrm{H}}^{min}$ \\
       & $(R_{\sun})$ & $(M_{\sun})$ & ($10^{51}$ erg) & $(10^{-2} M_{\sun})$
       & \multicolumn{2}{c}{(km\,s$^{-1}$)}\\
\noalign{\smallskip}
\hline
\noalign{\smallskip}
 1987A &  35  & 18   & 1.5   & 7.65 &  3000 & 600 \\
1999em & 500  & 19   & 1.3   & 3.6  &  660  & 700 \\
2000cb &  35  & 22.3 & 4.4   & 8.3  &  8400 & 440 \\
 2003Z & 230  & 14   & 0.245 & 0.63 &  535  & 360 \\
2004et & 1500 & 22.9 & 2.3   & 6.8  &  1000 & 300 \\
2005cs & 600  & 15.9 & 0.41  & 0.82 &  610  & 300 \\
2008in & 570  & 13.6 & 0.505 & 1.5  &  770  & 490 \\
2009kf & 2000 & 28.1 & 21.5  & 40.0 &  7700 & 410 \\
2012A  &  715 & 13.1 & 0.525 & 1.16 &  710  & 400 \\
\noalign{\smallskip}
\hline
\end{tabular}
\end{table}

We pursued three goals: to derive the basic parameters of SN~2012A, to probe
   the clumpiness using the H$\alpha$ and H$\beta$ lines, and to explore
   the possible effects of the ejecta clumpiness.
We find the ejecta mass $M_{env}=13.1\pm0.7~M_{\sun}$, the explosion energy
   $E=(5.25\pm0.6)\times10^{50}$ erg, the pre-SN radius 
   $R_0=715\pm100~R_{\sun}$, and the total $^{56}$Ni mass 
   $M_{\mathrm{Ni}}=0.012\pm0.002~M_{\sun}$.
We estimate the progenitor mass to be $15\pm1~M_{\sun}$.
The $^{56}$Ni mass estimate coincides with the value derived by \citet{TCF_13}.
Moreover, the ejecta mass and the explosion energy are close to those obtained
   by \citeauthor{TCF_13}.
However, our pre-SN radius is three times larger. 
We rule out significantly smaller radius, because the pre-SN radius is
   constrained by the initial luminosity peak; it cannot be reproduced
   in the case of a compact pre-SN star \citep{TCF_13}.

\begin{figure}[t]
   \includegraphics[width=\hsize, clip, trim=10 162 40 93]{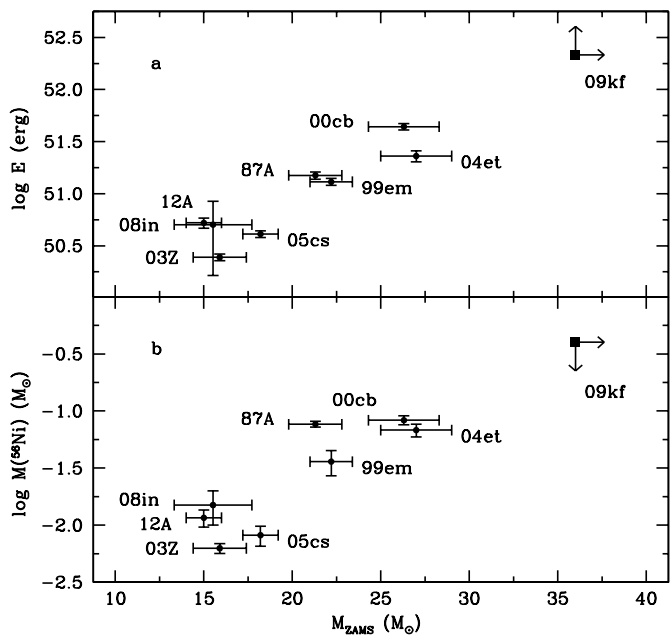}
   \caption{%
   Explosion energy (Panel \textbf{a}) and $^{56}$Ni mass (Panel \textbf{b})
      vs. hydrodynamic progenitor mass for SN~2012A and eight other
      core-collapse SNe \citep{UC_13}.
   The SN~2012A position on both scatter plots supports the ``explosion energy
      vs. progenitor mass'' and  ``$^{56}$Ni mass vs. progenitor mass'' 
      correlations.
   }
   \label{fig:ennims}
\end{figure}
With SN~2012A, we have now the parameters of nine SNe~IIP which are
   derived by the unique method of the hydrodynamic simulations (Table 1).
Among these objects two events, SN~1987A and SN~2000cb, are produced by the
   explosion of a blue supergiant and one peculiar event, SN~2009kf, has
   an anomalously high explosion energy.
The ejecta mass of SN~2012A turns out to be the smallest in this sample.
In the scatter plots of ``explosion energy vs. progenitor mass'' and
   ``$^{56}$Ni mass vs. progenitor mass'' (Fig.~\ref{fig:ennims}), the
   SN~2012A parameters fall into the bands occupied by other events.
In this regard SN~2012A is indeed the normal type IIP event.
 
The early spectra of SN~2012A show the H$\alpha$/H$\beta$ problem --
   a weak model H$\beta$ line for the model H$\alpha$ line consistent with
   observations -- recovered previously for SN~2008in \citep{UC_13, CU_14}.
The disparity is resolved in the same way as in the case of SN~2008in, i.e.,
   by invoking a clumpy structure of the external ejecta. 
The ejecta clumpiness is presumably produced during the shock wave propagation
   in the outer layers of a pre-SN which are associated with the density
   inversion and the vigorous convection in a RSG atmosphere. 
The mass of the clumpy external layers is estimated to be $\sim 0.07~M_{\sun}$.
This value is consistent with the mass of RSG layers of $0.01-0.2 M_{\sun}$
   above the density inversion for stars in the range of $10-20~M_{\sun}$
   \citep{Fad_12}.
Although proposed solution of the H$\alpha$/H$\beta$ problem seems to be
   reasonable, an independent decisive evidence should be found to confirm
   this conjecture.

Hydrodynamic simulations with the modified optimal model, which incorporates 
   a clumpiness of the outer $0.07~M_{\sun}$ layers into the radiation transfer,
   demonstrate that the most pronounced effect is the increase in the maximum
   velocity of the ejecta.
The physics behind this phenomenon is the clumpiness in the outer layers with
   the outermost smooth layer of low mass. 
The clumpiness favors the larger luminosity which results in the efficient
   radiative acceleration of the outer layers.

The effect of larger velocity in the clumpy outer ejecta is highly remarkable,
   because the low photospheric velocity at very early phase is a specific
   feature of hydrodynamic model of SNe~IIP with the low pre-SN mass
   \citep{UC_08}.
That was the reason why the ejecta mass and the explosion energy of the model
   have been pushed up in order to account for the observed large expansion
   velocity of the outer layers.
The increase of velocities of the outer ejecta in the clumpy model compared to
   the smooth one opens an interesting possibility to produce a hydrodynamic
   model of SNe~IIP with the lower pre-SN mass.
One can hope thus to resolve the mass problem, i.e., the conflict
   between the high hydrodynamic mass for SNe~IIP and the low mass based on
   the archival pre-explosion images.
The problem was first uncovered for SN~2005cs \citep{UC_08} when we compared
   the progenitor mass of $17-19~M_{\sun}$ obtained from the hydrodynamic
   modeling with the $7-12~M_{\sun}$ ZAMS mass recovered by \cite{MSD_05}
   from the archival images.
This disparity has emerged in several other cases as well.
For SN~2012A the hydrodynamic mass is slightly larger than the mass recovered
   from the pre-explosion image although they are consistent within errors
   \citep{TCF_13}.
The possibility to resolve the mass problem for SNe~IIP by invoking the ejecta
   clumpiness is promising and requires a separate study.

\begin{acknowledgements}
We thank Lina Tomasella for kindly sending us spectra of SN~2012A.
V.P.U. is grateful to Wolfgang Hillebrandt, Ewald M\"{u}ller, and Hans-Thomas
   Janka for hospitality during stay at the MPA.
V.P.U. is supported by Russian Scientific Foundation grant 14-12-00203.
\end{acknowledgements}



\end{document}